\documentclass[aps,prb,twocolumn,superscriptaddress,floatfix,nobibnotes]{revtex4-1}

\usepackage[utf8x]{inputenc}
\usepackage{amsmath,subfigure}
\usepackage[pdftex]{graphicx}
\usepackage{textcomp}

\usepackage{float} 
\floatstyle{boxed}

\usepackage{hyperref} 
\hypersetup{
    colorlinks=true,       
    linkcolor=red,          
    citecolor=blue,      
    filecolor=magenta,  
    urlcolor=blue,           
    urlbordercolor={0 1 1}}	

\begin{document}

\newfloat{copyrightfloat}{thp}{lop}
\begin{copyrightfloat}
\raggedright
The peer reviewed version of the following article has been published in final form at  J. Phys. Chem. B 2011, 115, 13513--13518, doi: \href{http://dx.doi.org/10.1021/jp2077215}{10.1021/jp2077215}.
\end{copyrightfloat}

\preprint{\today}

\title{Reversible and Irreversible Interactions of Poly(3-hexylthiophene) with Oxygen Studied by Spin-Sensitive Methods}

\author{Andreas Sperlich}
\email[]{sperlich@physik.uni-wuerzburg.de}
\author{Hannes Kraus}
\author{Carsten Deibel}
\affiliation{Experimental Physics VI, Julius-Maximilians-University of W\"urzburg, D-97074 W\"urzburg, Germany}
\author{Hubert Blok}
\author{Jan Schmidt}
\affiliation{Huygens Laboratory, Department of Molecular Physics, Leiden University, P.O. Box 9504, 2300 RA Leiden, The Netherlands}
\author{Vladimir Dyakonov}
\affiliation{Bavarian Centre for Applied Energy Research (ZAE Bayern), D-97074 W\"urzburg, Germany}
\affiliation{Experimental Physics VI, Julius-Maximilians-University of W\"urzburg, D-97074 W\"urzburg, Germany}


\begin{abstract}

Understanding of degradation mechanisms in polymer:fullerene bulk-heterojunctions on the microscopic level aimed at improving their intrinsic stability is crucial for the breakthrough of organic photovoltaics. These materials are vulnerable to exposure to light and/or oxygen, hence they involve electronic excitations. To unambiguously probe the excited states of various multiplicities and their reactions with oxygen, we applied combined magneto-optical methods based on multifrequency (9 and 275 GHz) electron paramagnetic resonance (EPR), photoluminescence (PL), and PL-detected magnetic resonance (PLDMR) to the conjugated polymer poly(3-hexylthiophene) (P3HT) and polymer:fullerene bulk heterojunctions (P3HT:PCBM; PCBM = [6,6]-phenyl-C61-butyric acid methyl ester). We identified two distinct photochemical reaction routes, one being fully reversible and related to the formation of polymer:oxygen charge transfer complexes, the other one, irreversible, being related to the formation of singlet oxygen under participation of bound triplet excitons on the polymer chain. With respect to the blends, we discuss the protective effect of the methanofullerenes on the conjugated polymer bypassing the triplet exciton generation.

\end{abstract}

\maketitle

\section{Introduction}\label{sec:introduction}

Conjugated polymers have received considerable attention as materials for the fabrication of organic electronics and photovoltaic devices due to their remarkable semiconducting and optoelectronic properties and the possibility to process devices from solution~\cite{Brabec:2008ji,Deibel:2010do}. For solar cells, being exposed to harsher conditions than many other device types, understanding the impact of different environmental influences is crucial to increasing the longevity of the cells~\cite{Jorgensen:2008dt}. The performance of a bulk heterojunction solar cell is determined not only by the energetic matching of donor and acceptor materials, but also by defect states hindering proper charge separation and extraction. Previous studies of oxygen-induced defect states in polymer devices show that charge carrier mobilities decrease by orders of magnitude upon exposure to air~\cite{Schafferhans:2008ho,Schafferhans:2010be}. Abdou et al.~\cite{Abdou:1997fo} noted that under illumination, poly(3-alkylthiophene) forms a weakly bound charge transfer (CT) complex with oxygen. Due to the low spectral resolution in \mbox{X-Band} electron paramagnetic resonance (EPR; 9.4~GHz), the assignment of the EPR signal to poly(3-hexylthiophene)$^{+}$ (P3HT$^{+}$) polarons or (P3HT:O$_2$)$^{-}_{\text{CT}}$ CT complexes remained elusive. Furthermore, L\"uer et al.~\cite{Luer:2004ky} found partially irreversible photoluminescence (PL) quenching of P3HT and attributed the origin to the quenching of singlet excitons with charged complexes. The chemical reactions behind the irreversible effects were examined by Hintz et al.~\cite{Hintz:2010eq}. They showed that photo-oxidation happens primarily on the alkyl side chains, forming peroxide species, which in turn destroy the thiophene unit, but the initial phototoxicity attacking the side chain is still presumed to be governed by triplet photosensitization~\cite{Evans:1990vw}.

Our intention is to complete this picture by using the possibilities of magnetic resonance to investigate the excitation pathway leading to the attack of the polymer side chain. EPR facilitates monitoring of polarons and charged complexes {\em in situ} avoiding the detour over electric contacts. By using high frequency (HF) EPR at 275~GHz, we were able to separate the spectra of P3HT$^{+}$ polarons and (P3HT:O$_2$)$^{-}_{\text{CT}}$ complexes and clarify that standard \mbox{X-Band} EPR shows the superposition of the two contributions as one line. Additionally, the measurement of PL intensity and its resonant change ($\Delta$PL) using the EPR extension, PL-detected magnetic resonance (PLDMR), allowed us to determine the degradation-related variations of the polymers' triplet yield.

\section{Experimental Details}
\subsection{Sample Preparation}
The P3HT polymer was purchased from Aldrich, and the [6,6]-phenyl-C61-butyric acid methyl ester (PCBM) was obtained from Solenne. No additional purification was performed. Sample preparation took place inside a nitrogen glovebox to avoid exposure to oxygen. The materials were dissolved in chlorobenzene with a concentration of 20~mg/mL. A 0.1~mL portion of the solution was poured into quartz EPR sample tubes, and the solvent was evaporated under rough vacuum (final pressure $3\times10^{-2}$~mbar), yielding a thick film on the inner sample tube wall. After drying, the sample was annealed for 10~min at 140~$^{\circ}$C. The P3HT sample tube stayed connected to the vacuum pump with an optional inlet valve for ambient air during the whole measurement cycle. The sample tube containing P3HT:PCBM (weight ratio 1:1) was sealed under vacuum with a blow torch.

\subsection{X-Band EPR}
To obtain EPR curves and the evolution transients, a modified \mbox{X-Band} spectrometer (Bruker 200D) was used (9.4~GHz~/~0.335~T). The sample was placed in a resonant cavity (Bruker ER4104OR) with optical access and cooled with a continuous flow helium cryostat (Oxford ESR 900) allowing a temperature range from 10~K to room temperature. Illumination was provided by a 532~nm DPSS laser arriving in the cavity on a sample-spot of roughly 4~mm in diameter with 80~mW, resulting in an intensity of $\sim$600~mW/cm$^{2}$. EPR was measured by using lock-in, with modulation of the external B-field as a reference. Simultaneously, the PL intensity of the sample was recorded with a silicon photodiode behind a 550--1000~nm filter. The time-dependent EPR measurements were performed at room temperature, and the magnetic field was set to a value corresponding to the \mbox{g-factor} g=2.0014 at the first derivative peak of the P3HT signal and stabilized by a field-frequency lock unit. The \mbox{g-factor} was calibrated for each measurement with the help of an NMR Gaussmeter (Bruker ER035M) and a microwave frequency counter (EIP 28b).

\subsection{PLDMR}
For PLDMR, the same setup as in \mbox{X-Band} EPR was used, with a frequency synthesizer (Wiltron 69137A) as the microwave source. The microwaves amplified by a solid state amplifier arrive in the cavity with a power of $\sim$60~mW. Instead of the microwave absorption, the variation of PL intensity ($\Delta$PL) due to resonant microwave irradiation was recorded. The measurements were performed with a lock-in amplifier, referenced by switching the microwave radiation on/off in the kHz range. All PLDMR measurements shown here were recorded at T=10~K.

\subsection{HF-EPR}
The HF-EPR measurements were performed at room temperature with the resonant conditions 275~GHz~/~9.8~T. Lock-in detection using field-modulation was applied in a similar way as in the \mbox{X-Band} setup. Further experimental details are given elsewhere \cite{Blok:2004hp,Blok2006phd}. For these measurements, P3HT was dissolved in dry toluene by stirring it at 60~$^{\circ}$C for 1~h. Then the solution was stirred at room temperature for one week, whereby the capping of the vial was removed to allow the toluene to evaporate. Care was taken to prevent exposure of the P3HT polymer to air by keeping the material in a glovebox during the sample preparation. The resulting P3HT powder was then inserted in a small diameter quartz tube and mounted into the 275~GHz cavity.

\section{Results}
\subsection{EPR}

\begin{figure}[htb]
\includegraphics{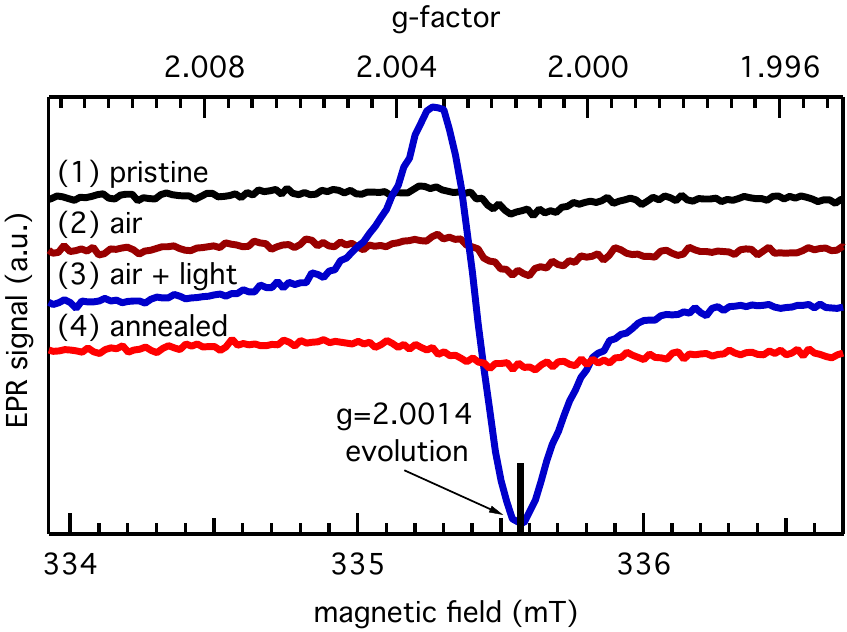}
\caption{\mbox{X-Band} EPR spectra of P3HT taken at 295~K without illumination: (1) as prepared in vacuum, (2) in air, (3) after air and light exposure (1~h), (4) after additional annealing for 10~min at 140~$^{\circ}$C in vacuum. The arrow points to the \mbox{g-factor} at which the evolution measurements were recorded (see \mbox{Figure \ref{fig:longterm})}.}
\label{fig:epr}
\end{figure}

\textbf{Figure \ref{fig:epr}} shows the influence of light and air on the \mbox{X-Band} EPR signal of P3HT recorded at room temperature without illumination. The initially very weak signal (1) is probably caused by residual impurities from the synthesis and can be ignored for further analysis. This signal is not caused by metallic catalysts, since those signals would usually be observed at much higher g-factor (low field region), and no additional signals have been found for these batches. Upon exposure to light {\em or} air (2) the signal rises very slightly. Only with the combined exposure to air and green laser (532~nm) illumination over 1~h can a substantial increase of the EPR signal be observed (3). This signal is persistent; the spectrum has been recorded in the dark after illumination. It can be completely removed (4) after an additional vacuum annealing step for 10~min at 140~$^{\circ}$C.

Abdou et al.~\cite{Abdou:1997fo} explained this rise of the EPR signal by the formation of oxygen-induced CT complexes. The spectral shift in EPR they showed can not be confirmed with the data presented here. Also, the reported intense signal of pure P3HT in vacuum is not in accordance with the vanishing weak initial signal in the pristine sample (see Figure \ref{fig:epr} (1)). Possibly, the P3HT presently available is of higher purity than was available in 1996. Due to the low spectral resolution of \mbox{X-Band} EPR, the assignment of the signal to CT complexes is rather ambiguous and needs further investigation.

\subsection{HF-EPR}

\begin{figure}[h]
\includegraphics{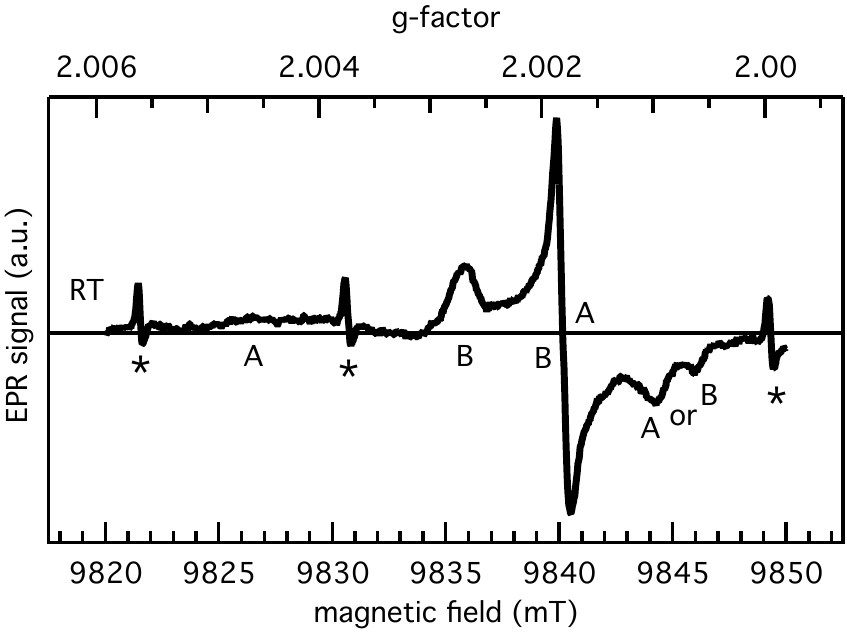}
\caption{The HF-EPR signal at 275~GHz of an illuminated P3HT sample at room temperature. Symbols A~and~B indicate the principal axes of the g-tensors of two paramagnetic species thought to be P3HT$^{+}$ and (P3HT:O$_{2}$)$^{-}_{\text{CT}}$ or vice versa. The asterisks (*) indicate signals of Mn$^{2+}$ ions serving as g-markers.}
\label{fig:hfepr}
\end{figure}

To clarify the assignment of the observed single EPR line at \mbox{X-Band} (9.4~GHz) to the proposed paramagnetic species, 275~GHz EPR experiments have been carried out at room temperature. \textbf{Figure \ref{fig:hfepr}} represents the result of the HF-EPR experiment on an illuminated P3HT sample. The two observed components are thought to be the EPR signals of the pair formed by P3HT$^{+}$ and the (P3HT:O$_{2}$)$^{-}_{\text{CT}}$ complex. From the comparison of the measured EPR curve with the simulated EPR spectrum of \mbox{g-anisotropy}-broadened EPR lines, it is concluded that the curve originates from at least two paramagnetic species A and B. In Figure \ref{fig:hfepr}, the principal axes of these species A and B are indicated. Due to the extremely small difference in the g-tensors, we have no means to attribute species A to P3HT$^{+}$ and B to (P3HT:O$_{2}$)$^{-}_{\text{CT}}$ or vice versa.

The P3HT sample was subsequently subjected to several treatments (similar to the experiments shown in Figure \ref{fig:epr}) at room temperature, and the effect was monitored on the 275~GHz EPR signals. First, the sample was kept in the dark in a gaseous helium atmosphere and the EPR signal was recorded. Then, still in the gaseous helium atmosphere, the sample was exposed to light from a Hg arc, and an identical EPR signal was observed as in Figure \ref{fig:hfepr}. Exposure to air without illumination did not affect the signal. A considerable increase was observed when the sample was exposed to light and air, and both signal contributions showed the same increase in intensity. This indicates that the formation of both P3HT$^{+}$ and (P3HT:O$_{2}$)$^{-}_{\text{CT}}$ is a coupled process, and that the stoichiometric ratio is 1. Finally, when subsequently keeping the sample in the dark, the signal did not change appreciably at room temperature.

The experiments show that it is the combined effect of light and air (oxygen) that causes the generation of the two paramagnetic species. In \mbox{X-Band} EPR, the superposition of both contributions is measured as one signal. 

\subsection{EPR of P3HT:PCBM Blends}
\begin{figure}[h]
\includegraphics{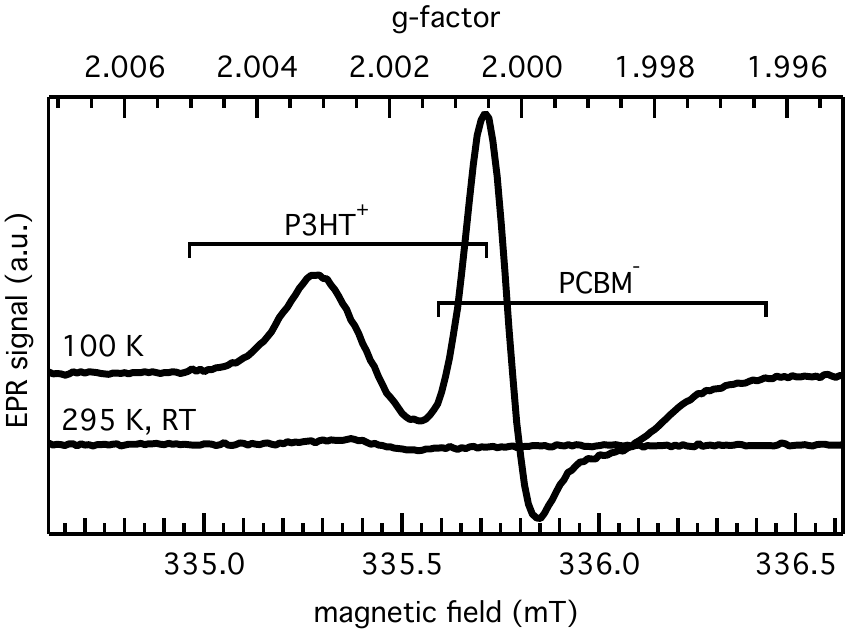}
\caption{\mbox{X-Band} EPR spectra of a P3HT:PCBM 1:1 blend taken at 100 and 295~K under illumination. The partially overlapping signals from P3HT$^{+}$ polarons and PCBM$^{-}$ anions at 100~K are marked. Almost no signal can be observed at 295~K.}
\label{fig:blend}
\end{figure}

\textbf{Figure \ref{fig:blend}} shows the \mbox{X-Band} EPR spectra of a P3HT:PCBM 1:1 blend at 100 and 295~K. At low temperatures, the typical well-known spectrum of the two overlapping signals of P3HT$^{+}$ polarons and PCBM$^{-}$ anions can be observed \cite{Ceuster:2001hc,Marumoto:2002ts}. However, at room temperature, only a marginal signal at g=2.002 can be detected, showing no illumination dependence, and can thus be assigned to residual impurities similar to the EPR signal observed in \mbox{Figure \ref{fig:epr}(1)}. In other words: The PCBM$^{-}$ anions and especially the P3HT$^{+}$ polarons can not be observed with standard EPR methods at room temperature. This is caused by the fast recombination of these charge carriers at room temperature~\cite{Marumoto:2002ts,Foertig:2009fw}. The fact that they can be observed as dedicated signals at lower temperatures stems from strong charge localization, i.e. trapping of charges in separate domains of the blended materials. This hinders recombination and therefore prolongs the charge carrier lifetime enormously, yielding the observed EPR spectra.

Thermal annealing of the blend samples does not change the observed lineshapes at low temperatures. However, the signal intensity decreases dramatically upon annealing due to small-scale phase separation and the resulting loss of interface surface. \cite{Savenije:2011dm,Poluektov:2010ie} The residual signal observed at room temperature (Figure \ref{fig:blend}) does not change substantially.

After comparison of the spectra shown in Figures 1--3 the following conclusion can be drawn: The oxygen-induced P3HT$^{+}$ polaron and (P3HT:O$_{2}$)$^{-}_{\text{CT}}$ complex demonstrate very slow recombination compared to P3HT:PCBM blends. In fact, the created paramagnetic species are persistent at room temperature and do not recombine readily after switching off the illumination. Thus at least one of the partners must be a trap state at room temperature hindering recombination.

\subsection{PL of P3HT under oxygen exposure}
For all samples studied, typical P3HT PL could be observed (see Supporting Information). Upon addition of PCBM or oxygen, these spectra were partly quenched, yet the spectral shape remained identical. This demonstrates that no other radiative recombination processes were enabled, and the singlet excitons responsible for PL of P3HT were unaffected. Thus the PL intensity is a proper scale to quantify singlet quenching in P3HT under oxygen exposure.

\subsection{EPR/PL Long-Term Evolution}

\begin{figure}[h]
\includegraphics{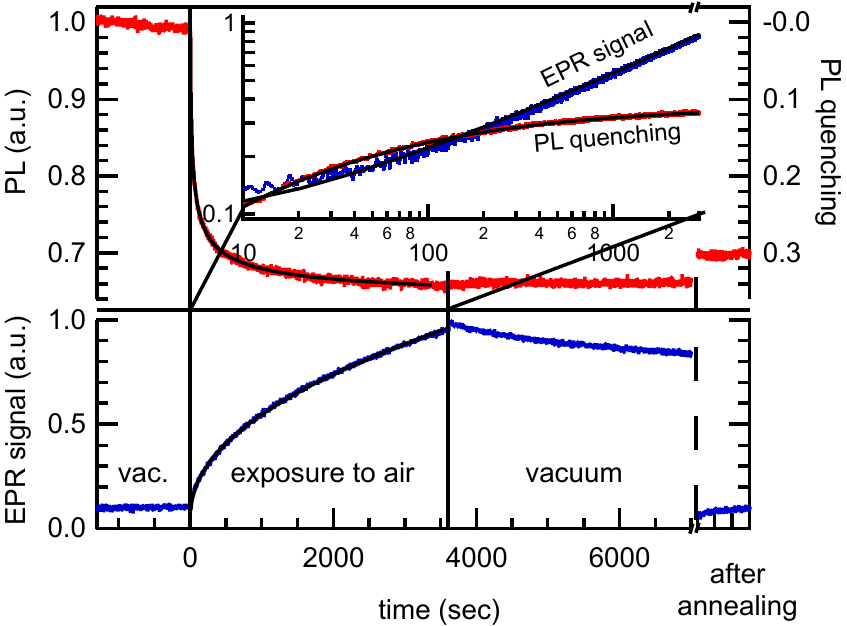}
\caption{The evolution of the EPR signal (g=2.0014 peak), the PL and the PL~quenching (=1-PL) at room temperature. Before 0~s, the sample was illuminated under vacuum. From that point on, air was introduced. This led to a continuing power law rise of the EPR signal and a decay of the PL. The inset shows a log--log plot of this time-range. From 3600~s on, the sample was evacuated, which led to slow decay of the EPR signal, but no change for the PL. After annealing, the PL recovered slightly, while the EPR signal was removed completely.}
\label{fig:longterm}
\end{figure}

After clarifying the origin of the EPR signals in oxygen-exposed P3HT, we studied the temporal evolution of this signal with \mbox{X-Band} EPR and simultaneously measured the PL intensity as shown in \textbf{Figure \ref{fig:longterm}}.

Before (t$<$0~s), the sample was in vacuum under constant illumination. PL and EPR signals changed only slightly in the time scale of hours, as long as the sample was exposed either to air (not shown) {\em or} light. We surmise that oxygen was desorbed into the P3HT, but without light any interaction was strongly suppressed, and the oxygen could be removed by evacuating the sample. This is consistent with the theoretical estimates of Lu and Meng~\cite{Lu:2007ig}, who postulate a highly oxygen-pressure-dependent hole-doping of P3HT in the dark. When using environmental air at atmospheric pressure (as we did), we would expect only marginal doping without light. 

The situation changes under concurrent light and air exposure from t$=$0~s on: the EPR signal rises, and the PL is quenched by 35~\% within 1~h. This evolution is independent of the gas diffusion into the sample bulk, which was confirmed by reversing the exposition order to first oxygen and then light (not shown). A diffusion time less than 1~s can be estimated from a layer thickness of $\approx1~$\textmu m and a diffusion constant for O$_{2}$ in P3HT films of 1.2~\textmu m$^{2}\text{s}^{-1}$ (from ref~\cite{Abdou:1997fo}) or even 0.15~\textmu m$^{2}\text{s}^{-1}$ (from ref~\cite{Luer:2004ky}).

From 3600~s (1~h) on, the sample was evacuated again. This did not result in a recovery of the PL, but the EPR signal slowly decreased within hours. Applying a thermal annealing step (140~$^{\circ}$C for 10~min) while still evacuating the sample removed the EPR signal, whereas the PL recovered only slightly. This means that the weak physical adsorption of molecular oxygen to P3HT is fully reversible when reducing the oxygen-pressure by evacuating. The annealing assists this process by creating additional thermal motion of the polymer. The fact that the PL remained quenched, although EPR measurements state that all CT complexes were removed, implies that another interaction process is responsible for the PL~quenching. We therefore conclude that the irreversible part of the PL~quenching is due to photo-oxidation of the polymer.

The inset in Figure \ref{fig:longterm} shows the time range of \mbox{0--3600~s} (light and air exposure) in a log--log plot for the EPR signal and the PL~quenching (=1-PL). The different curve shapes are obvious. The EPR enhancing and the PL~quenching occur at different time scales, supporting the assumption of two separate interaction processes between P3HT and oxygen. The EPR signal evolution can be fitted with a power law ($EPR~signal = c \cdot t^{\alpha}$) which is in accordance with former results on radiation-induced (in our case, laser light) defect production~\cite{Jonas:1997bg}. We found $\alpha = 0.477$ which should be dependent on the experimental conditions (partial O$_{2}$-pressure, light intensity). The PL~quenching was fitted using an equivalent formulation of the Hill equation \cite{Hill:1913tu,Weiss:1997wq}.

$PL~quenching = PL_{start}+ \dfrac{PL_{end} - PL_{start}}{\left(\tau / t \right)^{n}+1}$

We found $\tau = 21.7~s$ which is also dependent on experimental conditions. More interesting is the exponent $n=0.547$ describing the affinity of a macromolecule (e.g., P3HT) to undergo a chemical reaction with potential ligands (e.g., O$_{2}$). For $n>1$, a positively cooperative reaction is to be assumed, i.e., once one ligand is bound to the molecule, its affinity for other ligand molecules increases. For $n=1$, no change in affinity is to be observed. If $n<1$, the affinity of the molecule decreases during the ongoing interaction with further ligands. This seems to be the case for photo-oxidation of P3HT. Already degraded polymer-segments have a reduced affinity for further interaction with oxygen.

Regarding the reversibility of the oxygen doping processes, our results contradict the findings of Liao et. al.:~\cite{Liao:2008jd} They claim that the doping is reversible when facilitated with visible light, and only UV excitation leads to irreversible photo-oxidation. Our results concerning the PL~quenching show irreversible degradation already with green excitation light, however with a rather high light intensity of $\sim$600~mW/cm$^{2}$. L\"uer et al. \cite{Luer:2004ky} attributed the PL~quenching mainly to the formation of CT complexes, while we show that even after complete removal of these complexes, the PL remains quenched. Hence, only the slight reversible recovery of the PL after the thermal annealing might stem from PL~quenching due to CT complexes, while the irreversible part is photo-oxidation.

\subsection{PLDMR}
The open question is the connection between the reversible CT complex formation and irreversible photo-oxidation, and hence the destruction of the polymer. To answer it, we investigated the triplet generation properties of P3HT in relation to oxygen degradation. For P3HT it is well-known that, without an electron-accepting moiety, a significant part of generated singlet excitons undergoes an intersystem crossing (ISC) to the energetically favorable triplet state \cite{Dhoot:2002wu}. Other contributions to the triplet yield are the recombination of nongeminate polarons~\cite{Dyakonov:1998cd,Hachani:2008gm} or (in P3HT:fullerene blends) electron back transfer from the fullerene (lowest unoccupied molecular orbital) LUMO level if energetically possible~\cite{Liedtke:2011ci}.

\begin{figure}[h]
\includegraphics{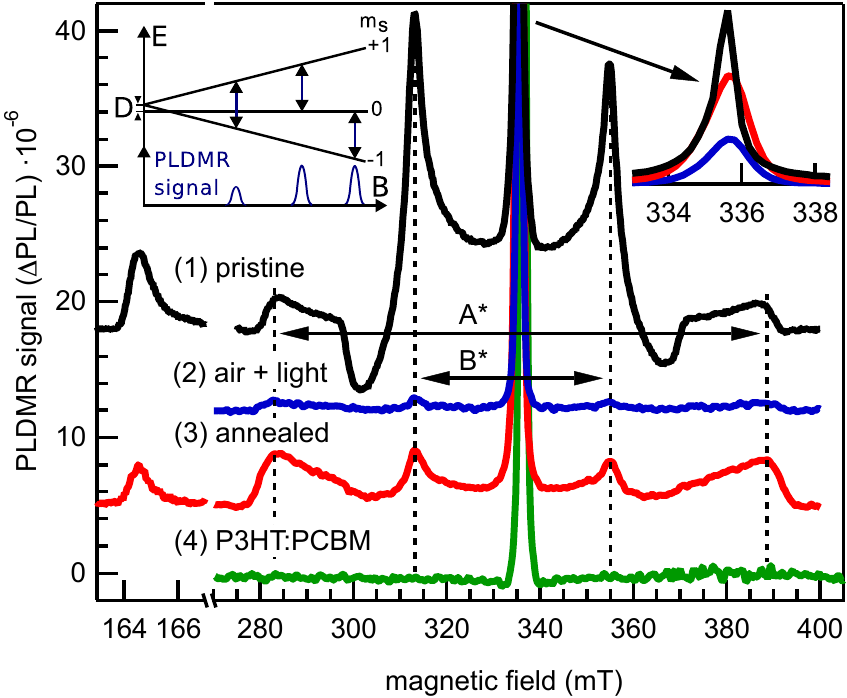}
\caption{PLDMR spectra taken at 10~K in vacuum. (1--3) P3HT: (1) as prepared, (2) after air and light exposure, (3) after additional annealing for 10~min at 140~$^{\circ}$C in vacuum. (4) P3HT:PCBM 1:1. The right inset shows a zoom on the central peak (P3HT 1--3) and the left inset shows the generation mechanism of the measured triplet pattern. Arrows A* and B* indicate possible triplet orientations.}
\label{fig:pldmr}
\end{figure}

\textbf{Figure \ref{fig:pldmr}} shows PLDMR spectra of P3HT and P3HT:PCBM 1:1. The $\Delta \text{m}_\text{s} = \pm 1$ and the first-order forbidden $\Delta \text{m}_\text{s} = \pm 2$ half field transition can be observed. The central peak in these spectra originates from nongeminate polaron pairs or can also be referred to as ``distant triplets'' with little or no zero-field splitting. The pristine curve (1) shows a distinct powder pattern observed for most thiophene-based polymers. Zero-field splitting D is responsible for the wing-like appearance of the powder pattern and is also correlated with the spatial extent of the corresponding triplet state. The different peaks result from the P3HT crystalline orientation in relation to the external magnetic field. We can distinguish between orientations A* and B*.

Note that the PLDMR signal is normalized to the photoluminescence ($\Delta \text{PL}/\text{PL}$), so the reduction of the total PL intensity has no influence on the overall signal height in PLDMR. 

The triplet exciton patterns in the pristine sample \mbox{(Figure \ref{fig:pldmr}(1))} are almost completely quenched after exposure to oxygen (2). The initially generated singlet excitons are separated before a possible ISC can occur, as oxygen is a very good electron acceptor~\cite{Kawaoka:1967ky}. Similar shape of the PLDMR spectrum was observed in P3HT:PCBM blends (4), where the CT from polymer to fullerene is the dominant process and therefore completely prevents ISC to the triplet state. Yet the PL~quenching is usually incomplete, leaving a small fraction ($<5$~\%) of the P3HT PL, enabling the application of PLDMR technique to blends. 

After removing the adsorbed oxygen by evacuating and annealing (see the EPR section), only chemically bound oxygen remains, which is responsible for the irreversible PL quenching. In the PLDMR signal, the reduction of the overall signal height (3) in comparison to the pristine PLDMR (1) is attributed to the triplet generation in P3HT being a strongly multimolecular process \cite{Partee:1999cq,Dyakonov:1997vx}. In a degraded polymer, the intact thiophene units have less other intact units in their vicinity, preventing triplet generation.


\section{Discussion}

\begin{figure}[h]
	\includegraphics{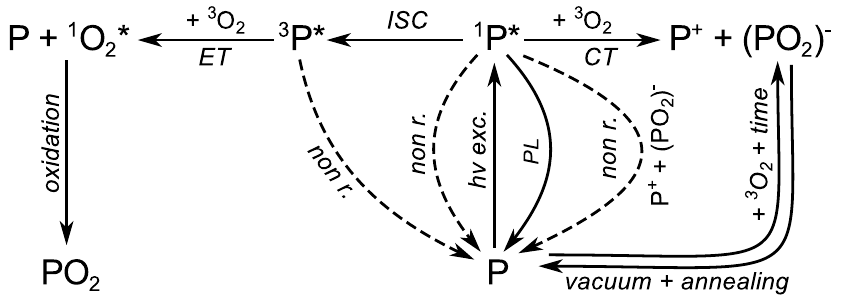}
	\caption{Oxygen related degradation processes in a conjugated polymer, P. Details are given in the text.}
  \label{fig:cartoon}
\end{figure}

\textbf{Figure \ref{fig:cartoon}} shows the proposed oxygen-related degradation process scheme in a conjugated polymer, P (e.g., P3HT). The polymer (P) can be excited by visible light ($h\nu~exc.$) yielding a singlet exciton ($^1$P$^*$). This exciton can recombine to the ground state by radiative ($PL$) and non-radiative ($non~r.$) processes. Another possibility is the relaxation of the singlet  via intersystem crossing ($ISC$) to the energetically favorable triplet state ($^3$P$^*$)~\cite{Dhoot:2002wu}, which can also be populated via alternative routes, e.g., via nongeminate polarons (not shown)~\cite{Dyakonov:1998cd,Hachani:2008gm}. Under certain conditions, the formation of triplets in polymer:fullerene blends is also possible as it depends on the relative energy positions of polymer triplet excitons and the fullerene LUMO level~\cite{Liedtke:2011ci}. Formed triplet excitons located on the polymer can then nonradiatively ($non~r.$) relax to the ground state. Additional reactions emerge upon introduction of molecular oxygen, which has a triplet ground state configuration ($^3$O$_2$).

The polymer triplet excitons can undergo energy transfer ($ET$) to the triplet oxygen to generate excited singlet oxygen ($^1$O$_2^*$)~\cite{Kawaoka:1967ky}. This is the reactive species for photo-oxidation of the polymer (depicted by the $oxidation$ to ``PO$_2$'') that irreversibly quenches PL and bleaches the UV--vis absorption by attacking the polymer~\cite{Hintz:2010eq,Evans:1990vw}.

Another possible mechanism for the production of singlet oxygen is presented in the literature: optically excited P3HT:oxygen complexes, which can in turn relax to singlet oxygen states~\cite{Scurlock:1989dl}. However, this process cannot be confirmed by the results presented in this work, since the PL~quenching saturates upon formation of CT complexes (see Figure \ref{fig:longterm}). Yet the absorption spectrum of CT complexes is unknown, and green laser illumination (532~nm) might not be suitable to initiate this reaction.

The other major reaction route shown is $CT$ from the polymer to oxygen to create positively charged polarons (P$^+$) and weakly bound CT complexes ((PO$_2$)$^-$). The creation of free charge carriers (P$^+$/ P$^-$) via singlet excitons is also possible, which then react with oxygen, yielding the same result (not shown)~\cite{Lu:2007ig}. This direct creation of free charge carriers is only relevant for high-energy photons (blue -- UV) as can be seen by the external quantum efficiency of a P3HT diode~\cite{Deibel:2010da}. The polarons and CT complexes formed give rise to the observed EPR signals in Figures \ref{fig:epr} and \ref{fig:hfepr} and can be removed slowly in $vacuum$ at room temperature (see Figure \ref{fig:longterm}: $t > 3600~s$) or rapidly by additional $annealing$ at 140~$^{\circ}$C (see also Figure \ref{fig:epr}). This reversible doping also occurs very slowly in the dark at room temperature as shown by EPR measurements and proposed by Lu and Meng~\cite{Lu:2007ig}, but proceeds orders of magnitudes faster upon excitation ($h\nu~exc.$) of the polymer.

The oxygen-induced doping in organic solar cells is responsible for a loss of short circuit current (J$_{\text{SC}}$) and reducing the overall mobility of charge carriers, while photo-oxidation of the devices also leads to a reduction in fill factor (FF) and open circuit voltage (V$_{\text{OC}}$)~\cite{Schafferhans:2010be}.

Additionally, the formation of polarons and CT complexes opens another nonradiative recombination channel ($non~r.$ / \mbox{(P$^+$ + (PO$_2$)$^-$))} for singlet excitons back to the ground state ($^1$P$^*$ $\rightarrow$ P)~\cite{Luer:2004ky}. This is the observed reversible part of the PL~quenching and blocks the less favorable $ISC$ to the triplet state. Thus the formation of polarons and CT complexes prevents further irreversible degradation by triplet-sensitized singlet oxygen. This can be observed in Figure \ref{fig:longterm}: with increasing EPR signal, the reduction of the PL intensity stops almost. This is backed up by the quenched triplet generation found in PLDMR (see Figure \ref{fig:pldmr}(2)). 

A similar protection effect through quenching of the triplet generation was seen in poly(phenylene vinylene) (PPV):PCBM blends, which are also more robust against oxygen exposure than pure PPV~\cite{Neugebauer:2000kq,Golovnin:2008gn}. This is in accordance with the presented PLDMR results, showing that in P3HT:PCBM blends, no triplets are generated (see Figure \ref{fig:pldmr}(4)) and thus would also prevent the creation of singlet oxygen over triplet photosensitization. However, this protection by quenching of the triplet production is not present in all polymer:fullerene blends and might be problematic, especially for new high-voltage blend combinations~\cite{Liedtke:2011ci}.

\section{Conclusion}

In conclusion, we demonstrated that at least two different mechanisms are responsible for the photodegradation of P3HT. With high-resolution EPR techniques, we have shown that the formation of CT complexes, previously proposed by Abdou et al., is fully reversible and is only partially responsible for the PL quenching.  Instead, the dominant PL quenching mechanism is the  irreversible photo-oxidation of the polymer by triplet-photosensitized singlet oxygen, as follows from the PLDMR. Important for the organic photovoltaic (OPV) applications is that the polymer degradation processes become significant only when it is exposed to light. Dark oxygen exposure can be reversed by annealing in a vacuum. These results are supported by the PLDMR measurements, which show a suppression of the triplet generation by polarons and CT complexes. In polymer--fullerene blends, a protective effect of CT complexes and fullerenes bypassing the formation of singlet oxygen and hence the degradation of the polymer can be anticipated.

\begin{acknowledgments}

The work at the University of W\"urzburg was supported by the BMBF OPV stability project under Contract No. 03SF0334F and the German Research Foundation, DFG, within the SPP 1355  ``Elementarprozesse der Organischen Photovoltaik''  (DY18/6-1 and 2). We thank Prof. E. J. J. Groenen from the Uni Leiden, Dr. T. J. Savenije from the TU Delft, H. Hintz from the Uni T\"ubingen, and Dr. J. Rauh (formerly Schafferhans) from the Uni W\"urzburg for fruitful discussions.
\end{acknowledgments}

\section*{Supporting Information Available}

PL spectra of P3HT, P3HT:PCBM, and P3HT under oxygen exposure.
This information is available free of charge via the Internet at  \href{http://pubs.acs.org/doi/suppl/10.1021/jp2077215}{http://pubs.acs.org}.

\bibliography{hex}

\end{document}